\documentclass[prd,twocolumn,superscriptaddress,showpacs,nofootinbib,preprintnumbers]{revtex4}

\usepackage{amsmath}
\usepackage{amsfonts}
\usepackage{graphicx}
\usepackage{dcolumn}

\def\be{\begin{equation}}
\def\ee{\end{equation}}
\def\ba{\begin{eqnarray}}
\def\ea{\end{eqnarray}}
\def\bs{\begin{subequations}}
\def\es{\end{subequations}}

\usepackage{color}

\begin{document}

\title{ Inflation driven by single geometric tachyon with D-brane orbiting \\around NS5-branes }
\author{Pyung Seong Kwon}
\email[E-mail me at: ]{bskwon@ks.ac.kr}
\affiliation{Department of Physics, Kyungsung University, Pusan
608-736,Korea}
\author{Gyeong Yun Jun}
\email[E-mail me at: ]{gyjun@ks.ac.kr}
\affiliation{Department of Physics, Kyungsung University, Pusan
608-736,Korea}
\author{Kamal L. Panigrahi}
\email[E-mail me at: ]{panigrahi@phy.iitkgp.ernet.in}
\affiliation{Department of Physics \& Meteorology and Centre for Theoretical Studies, Indian Institute of Technology Kharagpur- 721 302, India}
\author{M. Sami}
\email[E-mail me at: ]{sami@iucaa.ernet.in}
\affiliation{Center for Theoretical Physics, Jamia Millia Islamia, New Delhi-110092, India}

\begin{abstract}
We investigate models in which inflation is driven by a single
geometrical tachyon. We assume that the D-brane
 as a probe brane in the background of NS5-branes
has non-zero angular momentum which is shown to play similar role as
the number of the scalar fields of the assisted inflation. We
demonstrate that the angular momentum corrected  effective potential
allows to account for the observational constraint on COBE
normalization, spectral index $n_S$ and the tensor to scalar ratio
of perturbations consistent with WMAP seven years data.
\end{abstract}


\maketitle

\section{Introduction}
%%%%
The efforts to investigate  the time dependent backgrounds in string
theory have recently attracted attention in context with the study
of tachyon condensation \cite{senrev}. The effective field theoretic
set up to describe the dynamics of the rolling tachyon is provided
by Dirac-Born-Infeld (DBI) action \cite{DBI}. It was observed that
in the process of tachyon condensation, the unstable brane or the
brane-antibrane pair can decay to form a new stable D-brane.

It is interesting to note that the equation of state parameter of
the rolling tachyon field varies from zero to minus one which gave
rise to hope that the open string tachyon on the unstable D-brane
can play the role of inflaton \cite{tachinfl,KL} (see also Refs.in
~\cite{tachyonpapers} on the related issue). This idea was also
generalized to the radion field in the case of a brane moving
towards an antibrane, and vice versa \cite{BA}. But since the
effective potential of the rolling tachyon computed in the
perturbative string theoretic framework does not contain adjustable
parameters, it is not surprising that the model fails to be
compatible  with the requirement of slow-roll and COBE
normalization. Efforts were made to address these problems via
warped compactification \cite{warpcosm, GST} of the string theory.
Inspite of several attempts to overcome the problems in open string
tachyon cosmology, it seems unlikely that this tachyon field is
responsible for inflation. Attempts to assign the role of dark
matter fluid to rolling tachyon are also faced with a serious
problem of caustic formation which points towards the incompleteness
of the theory.

In string theory, however,there might be many interesting time
dependent backgrounds and one of such possibilities was recently
investigated in  \cite{kutasov}. In this scenario the motion of
D-brane, as a probe brane in the background of $k$ coincident
NS5-branes,  gives rise to an interesting dynamics which can be
mapped  to DBI action. Indeed, it was shown in the five-dimensional
brane world models (codimension-1 brane)  \cite{ekpark} and in the
$(p+3)$-dimensional string theory (codimension-2 brane)
\cite{pskwon} that the presence of
 NS-NS type brane is indispensable to
 obtain flat backgrounds on the transverse dimensions. This
 suggests that desired brane world models must involve the NS-brane as
 their background branes and the SM-branes(D-branes) are then placed near the background NS-branes.
These models, however, are faced with a instability problem of the
D-brane. It is known \cite{kutasov} that a D-brane propagating at
some distance from a stack of $k$ parallel NS5-branes becomes
unstable. The NS5-branes are much heavier than the D-branes in the
regime of small string coupling in this picture. Geometrically this
means that the NS5-branes form an infinite throat in space-time and
the string coupling increases as we move towards the bottom. Being
lighter, the probe brane is gravitationally pulled towards the
NS5-branes.

The D-brane preserves half of the supersymmetry which is different
from the other half preserved by the NS5-branes in Type-II theory.
Consequently, as the probe brane comes nearer the source brane,
the supersymmetry of the system is completely broken which gives
rise to a tachyonic degree of freedom on the D-brane. Indeed, the
radion becomes tachyonic in this case and then there is a map
between the tachyonic radion field living on the world volume of
the probe brane and the rolling tachyon associated with a non-BPS
D-brane. Thus the motion of the probe brane in the throat could be
described by the condensation of the tachyon. It is also possible
to study the motion of the probe brane in the background of a ring
of NS5-branes instead of coincident branes Refs.\cite{TW1,TW2}. It
is observed that the radion field becomes tachyonic when the probe
brane is confined to one dimensional motion inside the ring. The
condensation of the geometrical tachyon can play important role in
cosmology \cite{NS5cos,TWinf}.

Cosmological applications of these models with open string tachyon
have been studied in literature using BPS and non-BPS D-branes.
For instance in \cite{043}, an assisted inflation using open
string tachyons was studied, where $N$ open string tachyons are
allowed to roll simultaneously to obtain slow-roll inflation. In
\cite{044}, on the other hand, the authors studied rolling
geometrical tachyon induced on the $N$ D3-branes moving in the
vicinity of NS5-branes to show that this system coupled to gravity
gives the slow-roll assisted  inflation of the scalar field
theory. Further assisted inflation scenario from the rolling of
$N$ BPS D3-brane into the NS5-branes, on a transverse geometry of
$R^3 \times S^1$, coupled to four dimensional gravity has also
been studied in \cite{Panigrahi:2008kg}. In general,
tachyon-inflation models suffer from the large $\eta$-problem as
in the conventional models and are not favored for slow-roll
inflation. This difficulty may be avoided by allowing a large
number of tachyons to simultaneously roll down to assist the
inflation as mentioned above. Such an assisted inflation, however,
necessarily contains a large number of D3-brane in the theory and
consequently it is not adequate if we want to find a theory with a
single D3-brane. Thus in that case we may need to think about a
different type  of the theory which only involves a single
D3-brane but can satisfy the requirement of the slow-roll
inflation.

%%%%%
As mentioned above, in the configurations with D-brane(s) near
NS5-branes the supersymmetry of the system is completely broken
and this gives rise to a tachyonic degree of freedom on the
D-brane and it becomes a non-BPS D-brane. The D-brane eventually
falls into the fivebranes and decays into a pressureless fluid
called "tachyon matter". Such a decay of the D-brane, however, can
be avoided in special cases. In \cite{045}, it was demonstrated
that for certain values of energy and angular momentum the D-brane
orbits around the fivebranes, maintaining a fixed distance from
the fivebranes all the times and the decay of the D-brane is
suppressed. Indeed, in the case of nonzero angular momentum the
effective tachyon potential takes very different as compared to
the case of zero angular momentum \footnote{for further examples
see \cite{Nayak:2010kc}} and can give rise to viable inflationary
scenario.

In this paper we shall examine inflationary models in which the
inflation is driven by a single geometrical tachyon with an
assumption that the probe D-brane has non-zero angular momentum.

\section{D-brane dynamics near NS5-branes}
We begin our discussion by briefly reviewing the work presented in
\cite{kutasov} and \cite{045}. In the presence of {\it k} coincident
NS5-branes, the metric, dilaton and NS-NS 3-form fields are given by
\begin{eqnarray}
& &ds^2 =dx_{\mu}dx^{\mu} + H(x^n)dx^{m} dx^{m}\equiv G_{MN} dx^{M} dx^{N}, \nonumber\\
& &e^{2(\Phi-\Phi_0 )}=h(x^n),\\
& &H_{mnp}=-\epsilon^q _{mnp} \partial _q \Phi , \nonumber
\end{eqnarray}
where  $x^{\mu} (\mu = 0,1,...5)$ are the coordinates along the
world volume of the {\it k} coincident NS5-branes, while $x^{m}\;
(m=6,7,8,9)$ are the coordinates along the transverse dimensions.
Also $h(x^n)$ is a harmonic function,
\begin{equation}
 h=1 +{kl^2_s}/{r^2},
\end{equation}
where $r^2 =\sum_{n=6}^{9} x^m x_m$ and $l_s$ is the string length.

Let us consider a Dp-brane moving in the vicinity of the stack of
NS5-branes and stretched along the directions $(x_1 , ... x_p)$ with
$p\leq 5$. If we label the world volume of the D-brane by
$\xi^{\mu}$ , $\mu = 0,1,...p$. Then in the static gauge we have
$\xi^{\mu}=x^{\mu}$. The dynamics of the world volume fields of the
Dp-brane propagating in the above background fields is governed by
DBI (Dirac-Born-Infeld) action
\begin{equation}
S_p = - \tau_p \int d^{p+1}\xi e^{-(\Phi -\Phi_0)} \sqrt{-\det\mid G_{\mu\nu} +B_{\mu\nu}\mid }  ,
\end{equation}
where $G_{\mu \nu}$ and $B_{\mu \nu}$ are the pullbacks of $G_{MN}$ and $B_{MN}$:
\begin{equation}
G_{\mu \nu} = \frac{\partial x^{M}}{\partial \xi^{\mu}} \frac{\partial x^{N}}{\partial \xi^{\nu}}G_{MN} ,\;\;\;\;B_{\mu \nu} = \frac{\partial{x^M}}{\partial \xi^{\mu}} \frac{\partial{x^N}}{\partial \xi^{\nu}}B_{MN}\;.
\end{equation}
In Eq.(4) $x^{M} =(\xi^{\mu} , x^m) $, and $x^m$  represent the
position of the Dp-brane in the transverse space and they give rise
to world volume scalars $X^{m}(\xi ^{\mu})$. In this paper we shall
assume that the world volume components of the B-field vanish, i.e.,
$B_{\mu\nu}=0$ as it generally breaks the isotropy of the Dp-brane
world volume. In case of spatially homogeneous and isotropic
background,  $X^m = X^m (t)$. $G_{\mu \nu}$ reduces to
\begin{equation}
G_{\mu \nu} = \eta_{\mu \nu} + \delta^0 _{\mu} \delta^0 _{\nu}\; h(X^n)\dot{X}^{m}\dot{X}^m \;,
\end{equation}
and upon introducing polar coordinates, $X^6 = R \cos \theta $ and
$X^7 = R \sin\theta$, the action (3) takes the following form,
\begin{equation}
S_p  =-\tau_{p} \int dt \sqrt{h^{-1}(R)-(\dot{R}^2 + R^2 \dot{\theta}^2)}\;,
\end{equation}
where $h(R)$ is now $h(R)= 1+ kl^2/R^2$ and we have set the volume of the D$p$-brane equal to one.

The angular momentum and conserved energy following from (6) take the forms

\begin{equation}
L= \tau_{p} \frac{R^2 \dot{\theta}}{\sqrt{h^{-1}(R)-(\dot{R}^2 +R^2 \dot{\theta}^2)}} \;,
\end{equation}

\begin{equation}
E=\tau_p  \frac{1}{h\sqrt{h^{-1}(R)-(\dot{R}^2 +R^2 \dot{\theta}^2)}} \;.
\end{equation}
Solving these two equations in terms of $\dot{R}^2$ and
$\dot{\theta}^2$, one obtains
\begin{equation}
\dot{R}^2=\frac{1}{\varepsilon^2 h^2}[\varepsilon^2 h -(1+\frac{l^2}{R^2})] \;,
\end{equation}
 and
\begin{equation}
\dot{\theta}^2=\frac{1}{R^4h^2}\frac{l^2}{\varepsilon^2}\;,
\end{equation}
where $l$ and $\varepsilon$ are defined as $l=L/\tau_p$ and
$\varepsilon = E/\tau_p$, respectively. Note that the D-brane orbits
around the fivebranes, maintaining certain distance from the
fivebranes all the time, i.e., $\dot{R}=0$ provided the conditions
\begin{equation}
\varepsilon=1  \;\;\;\; \longleftrightarrow E=\tau_p  \;,
\end{equation}
and
\begin{equation}
l=\sqrt{k}l_s  \;\;\;\; \longleftrightarrow L=\sqrt{k}l_s \tau_p
\end{equation}
are satisfied.

We next consider an action,
\begin{equation}
\tilde{S_p}=- \tau_p \int dt \sqrt{1+ \frac{l^2}{R^2}}\sqrt{h^{-1}- \dot{R}^2}
\end{equation}
In \cite{045}, it was shown that the action $\tilde{S_p}$ in (13) is classically equivalent
 to $S_p$ in (6) as far as the radial motion is concerned. This in turn
  means that the tachyonic behavior described by $\tilde{S_p}$ is equivalent to  that of
  the original action $S_p$ as the (geometrical) tachyon $T$  is only a field redefinition of $R$:
\begin{equation}
dT=\sqrt{h(R)}dR
\end{equation}
The solution of (14) is \cite{kutasov}:
\begin{equation}
T(R)=\sqrt{kl^2_s + R^2} + \frac{1}{2}\sqrt{k}l_s \ln
\frac{\sqrt{kl_s^2 +R^2}-\sqrt{k}l_s}{\sqrt{kl_s^2
+R^2}+\sqrt{k}l_s}\;,
\end{equation}
and in terms of $T$, $\tilde{S}_p$ can be rewritten as
\begin{equation}
\tilde{S}_p = -\int dt \tilde{V}(T)\sqrt{1-\dot{T}^2} = \int dt L(t)\;,
\end{equation}
where $\tilde{V}(T)$ is given by
\begin{equation}
\tilde{V}(T)=\tau_p\frac{\sqrt{1+\frac{l^2}{R^2}}}{\sqrt{h(R(T))}} \;.
\end{equation}
It should be noted that $\tilde{V}(T)$ becomes flat for
$l=\sqrt{k}l_s$ and in this case the tachyon does not roll at all,
thereby indicating that the tachyonic
 degree of freedom induced on the D-brane disappears and the D-brane returns to the stable brane.

\section{Inflation from a single Geometrical tachyon}
In this section we shall examine the cosmological implications of
the geometrical tachyon in the inflationary universe. As mentioned
earlier, the assisted inflation was proposed to overcome the large
$\eta$ problem. The assisted inflation consists of introducing a
large number of D3-branes or equivalently a large number of scalar
fields. This idea was extended to inflationary model using
$N$-geometrical tachyons \cite{044}, where the slow roll could be
achieved by taking $N$ to be sufficiently large. In what follows,
however, we shall focus on  inflationary models which use a single
geometrical tachyon. It would be convenient to represent the DBI
action (16) in the following form,
\begin{equation}
\tilde{S}_p = -\int d^4 x \tilde{V}(\tilde{\Phi})\sqrt{-g}\sqrt{1+ \alpha^{\prime}(\partial_{\mu}\tilde{\Phi})^2}\;\; ,
\end{equation}
where $\tilde{\Phi}\equiv -T/\sqrt{\alpha^{\prime}}$, and $g_{\mu\nu}$, a four-dimensional
 metric introduced on the D3-brane is taken to be the FRW metric:
\begin{equation}
ds_4^2 = -dt^2 + a^2(t)[dr^2 + r^2d\Omega_2^2]\;.
\end{equation}
The action (18) taken with the Einstein-Hilbert action
\begin{equation}
S_{E}=\frac{M_{p}^2}{2}\int d^4 x \sqrt{-g}R \;,
\end{equation}
lead to the following evolution equations,
\begin{equation}
\ddot{\tilde{\Phi}} = -(1- \alpha^{\prime} \dot{\tilde{\Phi}}^2)(M_s^2 \frac{\tilde{V}_{,\tilde{\Phi}}}{\tilde{V}}+ 3H\dot{\tilde{\Phi}})\;\;,
\end{equation}
\begin{equation}
H^2 =\frac{8\pi G}{3}\frac{\tilde{V}(\tilde{\Phi})}{\sqrt{1-\alpha^{\prime}\dot{\tilde{\Phi}}^2  }}
\end{equation}
where $H(t)\equiv \dot{a}/a$ is the Hubble parameter. Upon using (slow-roll)
 condition $\alpha^{\prime}\dot{\tilde{\Phi}}^2 \ll 1$, (22) can be cast in the standard form
\begin{equation}
H^2 = \frac{8\pi G}{3} [V_{eff}(\psi)+ \frac{\dot{\psi}^2}{2}]\;\;,
\end{equation}
where $\psi$ is some normalized version of $\tilde{\Phi}$ and
$V_{eff}$ is the corresponding potential. In what follows we shall
present the explicit form of the effective potential.

We now turn our attention to inflation driven by geometrical
tachyon. The slow roll parameters are given by,
\begin{equation}
\epsilon = \frac{M_p^2}{2} \left(
\frac{V_{eff}^{\prime}}{V_{eff}}\right)^2 , \;\; \eta = M_{p}^2
\frac{V_{eff}^{\prime\prime}}{V_{eff}}\;.
\end{equation}
 In order to draw required number of e-folds, $\epsilon$ and $|\eta|$ have to be small. These
parameters are also related to the spectral index $n_s$ of the
scalar density fluctuations in the early universe as
\begin{equation}
n_s -1 \simeq -6\epsilon +2\eta\;,
\end{equation}
The observations from the CMB measurements \cite{040} require that
$n_s\simeq 0.95$ which also implies that $\epsilon \ll 1$, $|\eta|
\ll 1$. In addition to this, $\epsilon$ and $\eta$ are constrained
by the observation on
 amplitude of the primordial density perturbations,
\begin{equation}
\delta_s = \frac{1}{\pi \sqrt{75}}\frac{1}{M_p^3}\frac{V_{eff}^{3/2}}{V_{eff}^{\prime}}\lesssim 10^{-5}
\end{equation}
at horizon crossing. In the discussion to follow, we shall
investigate the particular cases of the effective potential imposing
conditions on the angular momentum $l$ of the D3 brane.

\subsection*{(a) case $l < \sqrt{k}l_s$}
 First we consider the case, $l<\sqrt{k}l_s$. From (17) we observe
 that $\tilde{V}(R)/\tau_{p}\rightarrow 1$ as $R\rightarrow\infty$, while $\tilde{V}(R)/\tau_{p}\rightarrow l/\sqrt{k}l_{s}$ as $R\rightarrow 0$. So, for $l<\sqrt{k}l_s$ the tachyon rolls from $R \sim \infty$ $(T\sim\infty)$ to $R \sim 0$ $(T \sim -\infty)$ (note that $\tilde{V}(R)$ is a monotonic function), and  the inflation is expected to occur at $R\gg \sqrt{k}l_s$.

To find  $V_{eff}(\psi)$, we rewrite Eq.(17) in the form,
\begin{equation}
\tilde{V}(\tilde{\Phi}) =
 \tau_3 [1+ \frac{(l^2 - kl_s^2)}{2\alpha^{\prime}\tilde{\Phi}^2} + O(\frac{1}{\tilde{\Phi}^4})]
\end{equation}
for $R \gg \sqrt{k}l_s \;$  (or $\tilde{\Phi}\gg 1)$. Substituting
(27) into (22), and assuming $\alpha^{\prime}\dot{\tilde{\Phi}}\ll
1$, we find that,
\begin{equation}
H^2 = \frac{8 \pi G}{3} [ \frac{M_s^4}{(2\pi)^{3}g_s} (1+  \frac{(l^2-kl_s^2)M_s^4}{2\Phi^2})+ \frac{1}{(2\pi)^{3}g_s}\frac{1}{2} \dot{\Phi}^2 ]\;\;,
\end{equation}
where $\Phi \equiv M_s \tilde{\Phi}$ and we have used $\tau_3 =
M_s^4/(2\pi)^3g_s$. Comparing (28) with (23), one obtains the
expression for the effective potential,
\begin{equation}
V_{eff}(\psi)= \frac{M_s^4}{(2\pi)^3 g_s}[1+ \frac{(l^2 -kl_s^2)M_s^4}{2(2\pi)^3g_s}\frac{1}{\psi^2} ]\;\;,
\end{equation}
where $\psi$ is defined by $\psi \equiv \Phi / \sqrt{(2\pi)^3g_s}$.
The slow-roll parameters are now given by
\begin{equation}
\epsilon \simeq XY^3\;, \;\;\;  \eta \simeq -6 XY^2\;,
\end{equation}
where
\begin{equation}
X = \frac{(2\pi)^3g_s}{2k}\frac{M_p^2}{M_s^2}\frac{1}{ \Delta_1(l) },\;\;\;
Y = \frac{kM_s^2}{\Phi^2}\Delta_1(l) ,
\end{equation}
and $\Delta_1(l) $ is defined as, $\Delta_1(l) =(kl_s^2 -
l^2)/kl_s^2$.

Let us next estimate the slow-roll parameters. We make the following
plausible choice\cite{044}
\begin{equation}
\frac{\Phi^2}{kM_s^2} \sim 10,\;\;\; k=1.
\end{equation}
 Since  $0
< \Delta_1(l) < 1$ for $l<\sqrt{k}l_s$, we see from (31) that
$Y(=6\epsilon/|\eta|) \sim 0.1$, if we take $l$ so as to satisfy
$\Delta_1(l) \sim O(1)$. For $X\sim 1$, we  have
\begin{equation}
\epsilon\sim 0.002,\;\;\; \eta \sim -0.01,
\end{equation}
which agrees with (25).
Also from (29) one finds that (26) can be written as
\begin{equation}
\delta_s = \frac{(2\pi)^2}{\sqrt{75}}(\frac{M_s}{M_p})\frac{1}{\sqrt{k\Delta_1(l) }}\frac{1}{XY^{3/2}} \sim 10^{-5},
\end{equation}
which then implies
\begin{equation}
\frac{M_s}{M_p} \sim 10^{-7}.
\end{equation}
It is interesting to note that the tensor to scalar ratio of
perturbations, $r=16 \epsilon \simeq 0.032$ is quite low in the
model.
 Finally, $X\sim 1$ together with (35) implies
\begin{equation}
g_s \sim 10^{-16},
\end{equation}
which is the realistic decoupling limit considered in "Little String
Theories" [LST] around one TeV \cite{014, 015}. However, (36) is not
compatible with the value of $M_s/M_p\sim 10^{-16}$ obtained in
these theories\cite{014, 015}.

\subsection*{(b)\; case $l> \sqrt{k}l_s$ with  inflation occurring at $R \ll \sqrt{k}l_s$}
Let us now consider, $l>\sqrt{k}l_s$. In this case the tachyon rolls
from $R\sim 0\; (T\sim -\infty)$ to $R\sim \infty\; (T\sim \infty)$.
Thus for $l>\sqrt{k}l_s$, inflation is expected to occur around
$R\ll \sqrt{k}l_s$ or possibly at $R\sim \sqrt{k}l_s$ if
$\sqrt{k}l_s$ is not so large.

In the region, $R\ll \sqrt{k}l_s$, $T(R)$ can be approximated by
$T(R)\sim \sqrt{k}l_s \ln R/\sqrt{k}l_s$;  from (17), we obtain
\begin{equation}
\tilde{V}(\tilde{\Phi})\cong \tau_{p}\frac{l}{\sqrt{k}l_s}[1-\frac{1}{2}\Delta_2(l) e^{\frac{2}{\sqrt{k}}\tilde{\Phi}}]\;,
\end{equation}
where $\Delta_2(l) \equiv (l^2 -\sqrt{k}l_s^2)/l^2$ and
$\tilde{\Phi} = T/\sqrt{\alpha^{\prime}}$. Substituting (37) into
(22) and comparing it with (23) one finds,
\begin{equation}
V_{eff}(\psi)= \frac{M_s^4}{(2\pi)^3 g_s}(\frac{l}{\sqrt{k}l_s})[1-\frac{1}{2}\Delta_2(l) e^{\beta(l)\psi/M_s}]\;.
\end{equation}
where $\beta(l)= 2(2\pi)^{3/2}g_s^{1/2}(l_s /\sqrt{k}l)^{1/2}$ and
$\psi= [(2\pi)^{-3}g_s^{-1}(l/\sqrt{k}l_s)]^{1/2}\Phi$ with $\Phi
\equiv M_s \tilde{\Phi}$.  In this  case the slow-roll parameters
assume the form,
\begin{equation}
\epsilon \simeq XY^2\;\;, \;\;\; \eta  \simeq -4XY\;\;,
\end{equation}
where
\begin{equation}
X=\frac{(2\pi)^3 g_s}{2k}\frac{M_p^2}{M_s^2}(\frac{\sqrt{k}l_s}{l})\;,\;\;\; Y=\Delta_2(l) e^{\beta(l)\psi/M_s}\;.
\end{equation}
Let us note that $0< \Delta_2(l) < 1$ for $l> \sqrt{k}l_s$ and
therefore $Y(=4\epsilon /|\eta|)$ $\to 0$ as $\beta(l)\psi/M_s \;
(\simeq \Phi/M_s)$  $\to -\infty$.

Now we estimate $\epsilon$ and $\eta$ with the following assumption,
\begin{equation}
\frac{R}{\sqrt{k}l_s} \sim e^{\frac{1}{\sqrt{k}} \frac{\Phi}{M_s}} \sim \frac{1}{\sqrt{10}},\;\;\; k=1,
\end{equation}
 Eqs.(32) and (40) imply that,  $Y(=4\epsilon/|\eta|)\sim 1/\sqrt{10}$
 if
 $\Delta_2(l) \sim O(1)$. Thus, if we set, $X \sim 2 \times 10^{-2}$,
the slow-roll parameters become
\begin{equation}
\epsilon \sim 0.002,\;\;\;   \eta \sim -0.024,
\end{equation}
which agrees with (25). Turing to (26), one can show that $\delta_s$
can be written as
\begin{equation}
\delta_s \simeq - \frac{1}{2\pi \sqrt{75}}(\frac{M_s}{M_p})\frac{1}{XY}
\end{equation}
in the present case, $\delta_s \sim 10^{-5}$ implies
\begin{equation}
\frac{M_s}{M_p}\sim 10^{-6}.
\end{equation}
 And $X\sim 10^{-2}$ together with (44) gives
\begin{equation}
g_s \sim  \hat{l} \times 10^{-16}\;,\;\;\;\;(\hat{l} \equiv l/\sqrt{k}l_s)\;.
\end{equation}
So, differently from the case (a), $g_s$ can not be determined in the case (b).

\vspace{1cm}

\subsection*{(c) case $l > \sqrt{k}l_s$ with inflation occurring at $R \sim \sqrt{k}l_s$ }
We finally consider the case,  $l > \sqrt{k}l_s$ again, but assume
that inflation occurs around $R \sim \sqrt{k}l_s$ this time. To find
$\tilde{V}(\tilde{\Phi})$, we first set
\begin{equation}
R = \sqrt{k}l_s + \eta, \;\;\; (\eta \ll \sqrt{k}l_s).
\end{equation}
Then we find from (15)
\begin{equation}
T\simeq \sqrt{2k}l_s [1+ \frac{1}{\sqrt{2}}\ln(\sqrt{2}-1)+ x -\frac{1}{4}x^2 ],
\end{equation}
where $x\equiv \eta / \sqrt{k}l_s$.  Using then Eq.(17), we find
that
\begin{widetext}
\begin{equation}
\tilde{V}(T)\simeq \tau_p (\frac{1+ (l/\sqrt{k}l_s)^2}{2})^{1/2}[1- \frac{1}{2}\Delta_3(l) x + \frac{1}{2}\Delta_3(l) \frac{[kl_s^2 + \frac{1}{4}(l^2 -kl_s^2)]}{(l^2 +kl_s^2)}x^2 ],
\end{equation}
\end{widetext}
where $\Delta_3(l)  \equiv (l^2 -kl_s^2)/(l^2 +kl_s^2)$.  Eq.(47)
can be solved by
\begin{equation}
x = C(\tilde{\Phi}) + \frac{1}{4}C^2(\tilde{\Phi})\;,
\end{equation}
where $C(\tilde{\Phi})$ is defined by
\begin{equation}
C(\tilde{\Phi}) = \frac{1}{\sqrt{2k}}(\tilde{\Phi} - \tilde{\Phi}_0)
\end{equation}
with $\tilde{\Phi}\equiv T/ \sqrt{\alpha^{\prime}}$ and $\tilde{\Phi}_0 \equiv \sqrt{2k}[1+ (1/\sqrt{2})\ln (\sqrt{2}-1)]$. So (48) can be rewritten as
\begin{widetext}
\begin{equation}
\tilde{V}(\tilde{\Phi}) = \tau_p (\frac{1 + \hat{l}^2}{2})^{1/2}[1- \frac{1}{2\sqrt{2k}}\Delta_3(l) (\tilde{\Phi} -\tilde{\Phi}_0) +\frac{1}{8k}\Delta_3(l) ( 1 + \hat{l}^2 )^{-1}(\tilde{\Phi} -\tilde{\Phi}_0 )^2  ].
\end{equation}
\end{widetext}
where $\hat{l}\equiv l/\sqrt{k}l_s$. Finally from (22) and (23) one finds
\begin{widetext}
\begin{eqnarray}
 V_{eff} = \frac{M_s^4}{(2\pi)^3
g_s}(\frac{1+\hat{l}^2}{2})^{1/2}[1 - \frac{1}{2\sqrt{2k}}\sqrt{
(2\pi)^3 g_s (\frac{2}{ 1+ \hat{l}^2} )^{1/2}} \;\Delta_3(l)
\frac{1}{M_s}( \psi -\psi_0 )  + \frac{(2\pi)^3 g_s}{4 \sqrt{2}k}
\frac{\Delta_3(l) }{(1+\hat{l}^2)^{3/2}}\frac{1}{M_s^2}(\psi -
\psi_0)^2 ]\:
\end{eqnarray}
\end{widetext}
where $\psi$ is defined by
$\psi =  [(2\pi)^3 g_s ]^{-1/2}[ (1+ \hat{l}^2)/2]^{1/4}\Phi$.

The slow-roll parameters and $\delta_s$ following from (52) are therefore given by
\begin{equation}
\epsilon \simeq \frac{(2\pi)^3 g_s}{8\sqrt{2}k} \frac{M_p^2}{M_s^2}\frac{\Delta^2_{3}(l)}{(1 + \hat{l}^2)^{1/2}}\;,\;\;\;
\eta\simeq \frac{(2\pi)^3g_s}{2\sqrt{2}k}\frac{M_p^2}{M_s^2}\frac{\Delta_3(l) }{(1 + \hat{l}^2 )^{3/2}}\;,
\end{equation}
and
\begin{equation}
\delta_s \simeq - \frac{2\sqrt{2k}}{\pi \sqrt{75}}\frac{1}{(2\pi)^3 g_s}\frac{M_s^3}{M_p^3}(\frac{1+ \hat{l}^2}{2}       )^{1/2}\Delta_{3}^{-1}(l).
\end{equation}
Also from (53) one finds
\begin{equation}
\frac{4\epsilon}{\eta} = (1 + \hat{l}^2)\Delta_3(l)  .
\end{equation}
To estimate the unknown parameters, first consider the case $l \sim
\sqrt{k}l_s$, where $\Delta_3(l) \rightarrow 0$ and $(1 +
\hat{l}^2)\rightarrow 2$. But in this case, $4\epsilon/\eta$ goes to
zero (also note that $\eta$ is positive in the case (c)) and the
condition (25) can never be satisfied.  We next consider, $l\gg
\sqrt{k}l_s$ which gives rise
 to $4\epsilon/\eta \simeq \hat{l}^2 \gg
 1 $, and as a result, $\eta$ is negligibly small as compared to $\epsilon$. Thus using Eq.(25) we
 have the following estimates,
\begin{equation}
\epsilon \sim 0.01 ,\;\;\; \eta \sim 0,
\end{equation}
and from $\epsilon \simeq 20 \times g_s
(M_p/M_s)^2 /\hat{l} \sim 10^{-2}$ and ~$ \delta_s \simeq -3 \times 10
^{-4}(M_s/ M_p)^3 \; \hat{l}/g_s \sim 2 \times 10^{-5}$
we obtain the ratio of scales,
\begin{equation}
\frac{M_p}{M_s} \sim 3 \times 10^4,~~
\hat{l}=\frac{l}{\sqrt{k}l_s}\sim  g_s \times 10^{12}.
\end{equation}
We again note that the tensor to scalar ratio of perturbations given
by $r \simeq 0.16$ in the present case is consistent with
observations. Eq.(58) may be thought  as the equation for the number
$N$ of the scalar fields appearing in the model of assisted
inflation \cite{044}. In \cite{044}, it was shown that the condition
$\delta_s \sim 2 \times 10^{-5}$ implies $N \sim (2\pi)^{3}g_s
\times 10^{10}$. In the case, $ l\gg \sqrt{k}l_s $ and the factor
$\hat{l}$ plays the similar role as the number of the scalar fields
(or the number of D3-branes) of the assisted inflation \cite{044}.

 Before closing this section, we wish to discuss two important
aspect of a realistic scenario that might come out of  model under
consideration. The first is related to back reaction on the D3-brane
worldvolume. We stress that the background NS5-branes are much
heavier than D3-branes as the mass of NS5-branes is of the order of
$\frac{1}{g^2_s}$, while that of D-branes goes like $\frac{1}{g_s}$
thereby in weak coupling limit, the NS5-branes are much heavier than
the D3-branes and hence the dynamics is almost entirely governed by
the gravitational attraction of the D3-branes into the core of
NS5-branes. Thus for sufficiently weak coupling, the back reaction
of these probe D3-branes can be ignored.

Let us now comment on the 4-dimensional physics of our model. The
typical relationship between 4-dimensional Planck mass and the
string mass is given by,
\begin{eqnarray}
\frac{M^2_p}{M^2_s} = v_0 / ((2\pi)^7 g^2_s)
\end{eqnarray}
where $v_0$ = volume/$(l_s)^6$ is the overall volume factor of the
compact CY3-fold. Presumably, to control stringy corrections, the
CY3 must have a large volume, $v_0 >> 1$. We consider the asymmetric
CY3 with the volume factor $v_0 \sim 10^3$ which is also preferred
by a small value of string coupling as discussed in
\cite{Panigrahi:2008kg}. For instance, in the case (c), $g_s$ can be
estimated as $\sim 10^{-5}$ from the value of $ M_p/M_s$ in (57).
 With that choice there
is no backreaction on CY3-fold and there is no more than one
D3-brane per unit CY measured in the units of string length.

\section{case (c) again}
In Sec.III we have considered three cases of inflationary models
where the D3-brane is moving around a stack of NS5-branes with
certain radial and also a nonzero angular velocity. Among these
models, of particular interest is the case (c) where the D3-brane is
moving away from the fivebranes and  inflation occurs at the radial
distance, $R\sim \sqrt{k}l_s$; the distance from the fivebranes as
small as the string length. In fact, if the D-brane has an energy of
the magnitude $\varepsilon\sim \hat{l}/\sqrt{2}$, $R\sim\sqrt{k}l_s$
becomes a turning point of the radial motion, and the situation is
that the inflation takes place at the minimum distance $R \sim
\sqrt{k}l_s$ from the fivebranes. For such a D-brane, the angular
velocity at $R\sim \sqrt{k}l_s$ is about $\dot{\theta}\sim
1/\sqrt{2k}l_s$
 (see (10)), and therefore the time it takes for the D-brane to make a one turn will be about
\begin{equation}
t \sim  l_s \times 10 ,
\end{equation}
which becomes $t \sim 10^{-38}$s using  Eq.(57).

Let us now estimate the time scale $H^{-1}$. From (23) and (52) we see that
\begin{equation}
H^{-1} \sim (2 \pi)^{3/2} \frac{M_p}{M_s^2}(\frac{g_s}{\hat{l}})^{1/2} \;,
\end{equation}
which upon using (57) gives
\begin{equation}
H^{-1} \sim \frac{M_p}{M_s^2} \times 10^{-5} \simeq 10^{-1} l_s \simeq (10^{15} GeV)^{-1} \;,
\end{equation}
and therefore the expansion time $\Delta t (\equiv 100 H^{-1}) $ should be about
\begin{equation}
\Delta t   \sim  l_s \times 10 .
\end{equation}
Eq.(61) shows that $\Delta t$ is of the same order as the time it
takes for the D-brane to circle around the fivebranes. So the
natural scenario associated with the case (c) may be described as
follows. A D-brane comes into a stack of NS5-branes with a certain
energy  $\varepsilon\sim \hat{l}/\sqrt{2}$ and a nonzero angular
momentum $l$. This D-brane approaches the fivebranes until the
radial distance becomes $R \sim \sqrt{k}l_s$, then it turns back
to $R \rightarrow \infty$ after circling the fivebranes about once
within the distance $R \sim \sqrt{k}l_s$ and during the
revolution, the inflation occurs on the D-brane. In this process,
the D-brane loses so much of its energy that it can not escape to
infinity. It gets captured by the fivebranes and starts orbiting
around them. Also the orbiting D-brane keeps losing its energy and
the angular momentum, and finally finds its stable state
corresponding to $\varepsilon \sim 1$ and $\hat{l} \sim 1$, for
which $\tilde{V}(T)$ becomes $\tilde{V}(T) \sim \tau_p$. This
scenario is quite distinguished from the conventional scenarios
based on the ordinary tachyon condensation.
%
%Figure1
%
%\begin{figure}[!h]
 %  \centering
 %  \includegraphics[scale=0.42]{pot.jpg}

  % \caption {With the potential $\tilde{V}(T)$ tachyon rolls
  % from the maximum $\tilde{V}(T)=\hat{l}\tau_p$ ($R=0$) to
  % the minimum $\tilde{V}(T) = \tau_p$ ($R = \infty$),while
  % for the conventional potential the tachyon rolls from the
  % maximum $V(T) = \tau_p$ $(R=0)$ to the minimum $V(T) = 0$ $(R=0)$.   \label{fig:jpg}}
%\end{figure}
%
In the case of the ordinary non-BPS D-brane ,the tachyon rolls from
the maximum of the tachyon potential $V(T) = \tau_p$ to the minimum
$V(T)=0$ and in this process the minimum of the potential describes
a tachyon vacuum corresponding to $\varepsilon = 0$, where there are
no physical open string states. In the case (c), however, the end of
the process corresponds to nearly a BPS D-brane (see \cite{046})
with $\varepsilon \sim 1$ and $\hat{l} \sim 1$, on to which the open
strings can end.

\section{Summary}
 In this letter, we have investigated inflationary models which use rolling
geometrical tachyon as inflaton induced on a D3-brane moving in the
vicinity of NS5-branes. These models are distinguished from those of
the assisted inflation in the sense that they include a single
geometrical tachyon to drive inflation. We have shown that the
configuration with a single D3-brane with large angular momentum is
effectively similar to the configuration with a large number of
coincidence D3-branes as far as an inflation driven by rolling
geometrical tachyon(s) is concerned and this enables us to obtain a
scenario which can satisfy the requirement of slow-roll without
introducing a large number of D3-branes

Allowing for nonzero angular momentum $l$ configurations, leads to
an important feature in the tachyon potential, $\tilde{V}(T)=\tau_p
\sqrt{1+(l^2/R^2)}/\sqrt{h(R(T))}$. The potential $\tilde{V}(T) $
crucially differs from the conventional geometric tachyon potential
$V(T)$: It asymptotically approaches the value $\hat{l}\tau_p$ as $R
\rightarrow 0$ where $\hat{l} \equiv l/\sqrt{k}l_s $, while it
approaches $\tau_p$ as $R \rightarrow \infty$. Hence in case,
$l>\sqrt{k}l_s $ (or $\hat{l}>1)$ ,the tachyon rolls from the
maximum $\tilde{V}(T)=\hat{l}\tau_p$ ($R=0$) to the minimum
$\tilde{V}(T) = \tau_p$ ($R = \infty$).  As a result, inflation
occurs around $R \sim 0$, and thereafter the D-brane moves away and
finds its stable orbit at  $R \sim \infty$. We have shown that
amongst the various cases discussed in the present paper, the most
interesting possibility corresponds to $l
> \sqrt{k}l_s$.  In this case,  inflation  occurs  around $R \sim \sqrt{k}l_s $
and the end of the tachyon condensation corresponds to a BPS D-brane
where the open strings can end. This is quite different from the
conventional  geometrical tachyon scenario where tachyon rolls from
the maximum of its potential, $V(T)=\tau_p$ ($R = \infty$) to the
minimum, $V(T)=0$ ($R =0$. As a consequence, the D-brane finally
decays into a tachyon vacuum as it approaches $R =0$ where there are
no physical open string states.

It is interesting to note that in the scenario discussed here, the
estimate of expansion time $H^{-1}$ corresponds to GUT scale which
is, of course, related to the assumptions we made in the text. For
instance, in the case (c), namely, $l
>\sqrt{k} l_s$, we assumed that the inflation occurs at the turning
point $R \sim \sqrt{k}l_s$ of the radial motion of the D-brane.
However, it is also possible to consider other configurations with
different turning points allowing us  to generate different values
of the expansion scale.
\\
\\

\begin{acknowledgements}
P. S. Kwon was supported by the grant of Kyungsung University.
\end{acknowledgements}

%%%%%%%

\end{document}